\documentclass[journal,comsoc]{IEEEtran}

\ifCLASSINFOpdf
\else
\fi

\usepackage[cmex10]{amsmath}
\usepackage{amssymb}
\usepackage{siunitx}
\usepackage{caption}
\usepackage{algpseudocode}
\usepackage{algorithm}
\algnewcommand{\Initialization}[1]{%
  \State \textbf{initialization:}
  \Statex \hspace*{\algorithmicindent}\parbox[t]{.8\linewidth}{\raggedright #1}
}

\algnewcommand{\Rep}[1]{%
  \State \textbf{repeat:}
  \Statex \hspace*{\algorithmicindent}\parbox[t]{.8\linewidth}{\raggedright #1}
}

\usepackage{tikz}
\usetikzlibrary{shapes,arrows,fit,positioning,calc}
\usetikzlibrary{plotmarks}
\usetikzlibrary{decorations.pathreplacing}
\usetikzlibrary{patterns}
\usetikzlibrary{arrows,automata}

\usepackage{pgfplots}
\pgfplotsset{compat=newest}

\begin{document}

%
\title{Branch-and-Bound Precoding for Multiuser MIMO Systems with 1-Bit Quantization}

\author{Lukas~T.~N.~Landau,~\IEEEmembership{Member,~IEEE,}
        and~Rodrigo C.\ de Lamare,~\IEEEmembership{Senior~Member,~IEEE}\vspace{-1em}				
\thanks{The authors would like to thank Johannes Israel from the Institute of Numerical Mathematics, TU Dresden for the introduction of the branch-and-bound method and for verifying the notation.}
\thanks{The authors are with Centro de Estudos em Telecomunica\c{c}\~{o}es Pontif\'{i}cia Universidade Cat\'{o}lica do Rio de Janeiro, Rio de Janeiro CEP 22453-900, Brazil, (email: \{lukas.landau, delamare\}@cetuc.puc-rio.br).} }
\maketitle


\begin{abstract}
Multiple-antenna systems is a key technique to serve multiple users in future wireless systems.
For low energy consumption and hardware complexity we first consider transmit symbols with constant magnitude and then 1-bit digital-to-analog converters. We propose precoding designs which maximize the minimum distance to the decision threshold at the receiver.
The precoding design with 1-bit DAC corresponds to a discrete optimization problem, which we solve exactly with a branch-and-bound strategy. We alternatively present an approximation based on relaxation. Our results show that the proposed branch-and-bound approach has polynomial complexity. The proposed methods outperform existing precoding methods with 1-bit DAC in terms of uncoded bit error rate and sum-rate. The performance loss in comparison to infinite DAC resolution is small.
\end{abstract}



%

\vspace{-1em}
\section{Introduction}
Low peak-to-average ratio is essential for the use of cheap and efficient power amplifiers, which are key in
multiuser multiple-input multiple-output (MIMO) communication systems \cite{Spencer_2004}.
Especially in short range wireless communications with a low path loss, also the converters become an important factor of system cost.
It is known that digital-to-analog converters (DACs) have a lower energy consumption in comparison to analog-to-digital converters (ADCs) with the same clock speed and resolution. For example, when considering the converter pair presented in \cite{Naber_1989} and assuming that the energy consumption of a DAC or ADC doubles with every additional bit of resolution, the DAC consumes only $30\%$ of the energy consumption of the ADC with the same parameters. 
For this reason, the DAC is often neglected in the optimization of communications systems.
However, when considering the recent trends in multiuser MIMO communication, the number of antennas at the base station (BS) is commonly larger than the total number of receive antennas, which guides our attention to the DACs at the BS.
In this regard, DACs with 1-bit resolution are promising, where we assume that the pulse shaping can be efficiently performed in the analog domain.    
For some applications, e.g., internet of things, where the receivers should have a low-complexity and limited energy budget it is reasonable to also consider 1-bit quantization at the receivers.
In this context, MIMO communication systems with 1-bit quantization have received increased attention, where the design of the precoder is a particular problem.
In this regard, linear precoding strategies \cite{Saxena_2016} and \cite{JacobssonDCGS16a}, such as maximal-ratio transmission and zero-forcing (ZF) methods, followed by quantization have been studied. At the same time a nonlinear approach has been reported in \cite{Jedda_2016}, where the precoding vector is obtained based on an optimization method. Another nonlinear precoder has been presented in \cite{Tirkonnen_2017} whose computational complexity is only linear with the number of antennas.

In this work, we develop a precoding design that maximizes the minimum distance to the decision thresholds at the receivers which is promising in terms of bit error rate (BER). First we propose a precoding design with constant magnitude transmit symbols, termed phase-only precoding (PoP), and then we consider 1-bit DACs, which then corresponds to a scaled version of an integer linear program. We optimally solve this non convex problem by a branch-and-bound method, which has been lately considered for discrete receive beamforming \cite{Israel_2015_Letter}.
The bounding step relies on a relaxed problem which is a linear program (LP). This relaxation is also used to approximate the optimal precoder as an alternative solution.

The paper is organized as follows: Section~\ref{sec:system_model} describes the system model, whereas Section~\ref{sec:PoP} describes the proposed precoding design with constant magnitude transmit symbols. In Section~\ref{sec:BB_precoding}, we describe the proposed precoding algorithms with 1-bit DACs. Section~\ref{sec:numerical_results} presents and discusses numerical results, while Section VI gives the conclusions.

We use the notation $P(\boldsymbol{y}_k \vert \boldsymbol{s}_k)=P(\mathbf{y}_k=\boldsymbol{y}_k \vert \mathbf{s}_k=\boldsymbol{s}_k)$, where the random quantities are upright letter and the realizations italic.

%

\section{System Model}
\label{sec:system_model}
\begin{figure*}[h]
\begin{center}
\captionsetup{justification=centering}
\includegraphics{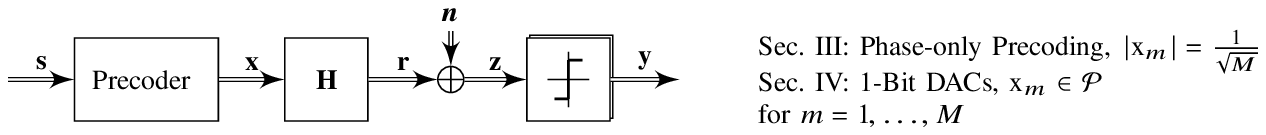}
\caption{MIMO system model}
\label{fig:system_model}       
\vspace{-0.75em}
\end{center}
\end{figure*}
A multiuser MIMO downlink is considered where the BS has $M$ transmit antennas which communicates with $K$ users, each having $L$ antennas.
The transmit vector is denoted by $\mathbf{x}=[\mathrm{x}_1,\ldots,\mathrm{x}_M]^T$.
The vector of transmit symbols of length $K L$ is denoted by $\mathbf{s}=[\mathbf{s}_1^T,\ldots,\mathbf{s}_K^T]^T$ with $\mathbf{s}_k=[\mathrm{s}_{k,1},\ldots,\mathrm{s}_{k,L}]^T$ and $\mathrm{s}_{k,l} \in \left\{  1+j,1-j,-1+j,-1-j  \right\}$ for $k=1,\ldots,K$ and $l=1,\ldots,L$. It is assumed a frequency flat fading channel described with zero-mean independent and identically distributed (i.i.d.) complex Gaussian random variables $\mathrm{h}_{k,l,m}$, where $k$, $l$ and $m$ denote the index of the user, the receive antenna and the transmit antenna, respectively.
After matched filtering at the transmitter and the receiver the received sample at the $l$th antenna of the $k$th user is described by\looseness-1 
\begin{align}
\label{eq:received_sample}
\mathrm{z}_{k,l} & =  \mathrm{r}_{k,l} +  \mathrm{n}_{k,l}  = \sum_{m=1}^{M}     \mathrm{h}_{k,l,m } \ \mathrm{x}_m + \mathrm{n}_{k,l} \textrm{,}
\end{align}
where $\mathrm{n}_{k,l}$ is a zero-mean i.i.d.\ complex Gaussian random variable with variance $\sigma_n^2$ representing thermal noise.
The received signals are applied to hard decision detectors or equivalently 1-bit ADCs described by $\mathrm{y}_{k,l}=Q(\mathrm{z}_{k,l})= \mathrm{sgn}(\mathrm{Re}\left\{\mathrm{z}_{k,l}\right\}) +j \ \mathrm{sgn}(\mathrm{Im}\left\{\mathrm{z}_{k,l}\right\})$, such that ${\mathrm{y}}_{k,l} \in \left\{ 1+j,1-j,-1+j,-1-j \right\}$, where $Q(\cdot)$ denotes the 1-bit quantization. 
By using vector notation the received $K L$ samples can be expressed as
\begin{align}
\mathbf{y}=Q(\mathbf{z})= Q(\mathbf{r} + \mathbf{n}) =Q(  \mathbf{H} \mathbf{x} + \mathbf{n}) \textrm{,}
\end{align}
where $\mathbf{H}=[\mathbf{H}_1^T,\ldots,\mathbf{H}_K^T]$ is the channel matrix with dimensions $K L \times M$ which consists of the matrices $\mathbf{H}_k$ each having dimensions $L \times M$. The compact system model is illustrated in Fig.~\ref{fig:system_model}.       
The receive vector has the structure $\mathbf{y}=[\mathbf{y}_1^T,\ldots,\mathbf{y}_K^T]^T$ accordingly. 
The following describes, how to choose the transmit vector $\mathbf{x}$ such that transmit symbols $\mathbf{s}$ can be appropriately detected.

\section{Proposed Phase-only Precoding (PoP)}
\label{sec:PoP}
Instead of considering the BER or the achievable rate as the objective function, we consider a design criterion which has been proposed before in \cite{Landau_SCC2013} in the context of intersymbol interference, and which has been used later also in \cite{Mo_2015}, \cite{Gokceoglu_2016b}.
The design criterion implies that $\mathbf{x}$ is chosen such that the minimum distance to the decision threshold, denoted by $\epsilon$, is maximized, which provides robustness against perturbations of the received signals. In the proposed Phase-only Precoder (PoP) it is considered that each $x_m$ has constant magnitude.
By using instead the corresponding inequality, the optimization problem can be cast as: \looseness-1
\begin{align}
\label{eq:PoP}
\mathbf{x}_{\textrm{opt}} = &  \arg\min_{\mathbf{x}, \mathrm{\epsilon}}  -\mathrm{\epsilon} \\
& \textrm{s.t. }     \mathrm{Re} \left\{   \mathrm{diag} \left\{ \mathbf{s} \right\}  \right\}    \mathrm{Re} \left\{ \mathbf{H} \mathbf{x}  \right\}   \geq  \epsilon \boldsymbol{1}_{K L} \textrm{,} \notag \\
& \ \ \ \            \mathrm{Im} \left\{   \mathrm{diag} \left\{ \mathbf{s} \right\}  \right\}    \mathrm{Im} \left\{ \mathbf{H} \mathbf{x}  \right\}   \geq  \epsilon \boldsymbol{1}_{K L} \textrm{,} \notag \\
& \ \ \ \      \left|  \mathrm{x}_m \right| \leq   1/\sqrt{M}, \ \ \textrm{for } m=1,\ldots,M   \textrm{,}   \notag 
\end{align}
where the negation is applied to obtain a minimization problem.
If needed, the equality $\left|  \mathrm{x}_m \right| =   1/\sqrt{M}$ is subsequently introduced by scaling the entries of $\mathbf{x}_{\textrm{opt}}$.

\section{Proposed Precoding with 1-Bit DAC}
\label{sec:BB_precoding}
In this section, we propose a precoder designs using the objective from Section~\ref{sec:PoP} but with DACs having 1-bit resolution where we implicitly assume analog pulse shaping. With this, the optimization problem changes to
\begin{align}
\label{eq:optimization_problem_precoder}
\mathbf{x}_{\textrm{opt}} = &  \arg\min_{\mathbf{x}, \epsilon}  -\epsilon \\
& \textrm{s.t. }     \mathrm{Re} \left\{   \mathrm{diag} \left\{ \mathbf{s} \right\}  \right\}    \mathrm{Re} \left\{ \mathbf{H} \mathbf{x}  \right\}   \geq  \epsilon \boldsymbol{1}_{K L} \textrm{,} \notag \\
& \ \ \ \          \mathrm{Im} \left\{   \mathrm{diag} \left\{ \mathbf{s} \right\}  \right\}    \mathrm{Im} \left\{ \mathbf{H} \mathbf{x}  \right\}   \geq  \epsilon \boldsymbol{1}_{K L} \textrm{,} \notag \\
& \ \ \ \      \mathbf{x} \in \mathcal{P}^M \textrm{,}   \notag 
\end{align}
such that the elements of $\mathbf{x}$, termed $\mathrm{x}_m$ for $m=1,\ldots,M$ are taken from the set \newline $\mathcal{P}:= \left\{   e^{j \frac{\pi}{4}}/ \sqrt{M}, e^{j \frac{ 3\pi}{ 4 }}/ \sqrt{M}, e^{j \frac{ 5\pi}{ 4 }}/ \sqrt{M}, e^{j \frac{ 7\pi}{ 4 }}/ \sqrt{M}     \right\}$. Due to the constraint the optimization problem is non-convex.

\subsection{Proposed Approximate 1-Bit Precoder (1-Bit Approx.)}
\label{sec:lower-bounding}
The first step in the algorithm corresponds to a lower-bounding of the objective by relaxation.
Relaxing the input constraint in \eqref{eq:optimization_problem_precoder} yields the LP: 
\begin{align}
\label{eq:lower_bound}
\mathbf{x}_{\textrm{lb}}=&  \arg\min_{\mathbf{x}, \epsilon}  -\epsilon \\
& \textrm{s.t. }     \mathrm{Re} \left\{   \mathrm{diag} \left\{ \mathbf{s} \right\}  \right\}    \mathrm{Re} \left\{ \mathbf{H} \mathbf{x}  \right\}   \geq  \epsilon \boldsymbol{1}_{K L} \textrm{,} \notag \\
& \ \ \ \            \mathrm{Im} \left\{   \mathrm{diag} \left\{ \mathbf{s} \right\}  \right\}    \mathrm{Im} \left\{ \mathbf{H} \mathbf{x}  \right\}   \geq  \epsilon \boldsymbol{1}_{K L} \textrm{,} \notag \\
&\ \ \ \   \left| \mathrm{Re} \left\{{\mathrm{x}}_m \right\}\right| \leq  1/\sqrt{2M}  ,     \notag \\
& \ \ \ \   \left| \mathrm{Im} \left\{{\mathrm{x}}_m \right\}\right| \leq   1/\sqrt{2M} , \ \ \textrm{for } m=1,\ldots,M  \textrm{.}   \notag
\end{align}
The optimal value of \eqref{eq:lower_bound} is always smaller than or equal to the optimal value of \eqref{eq:optimization_problem_precoder}. A valid solution in terms of $\mathbf{x} \in \mathcal{P}^M$ can be obtained by mapping the solution of \eqref{eq:lower_bound} to the discrete input vector with the smallest Euclidean distance. This corresponding design is the proposed approximate 1-bit precoder. The resulting $-\epsilon$ is an upperbound of the optimal value of \eqref{eq:optimization_problem_precoder}.

\subsection{Concept of Branch-and-Bound Precoding}
We consider a precoder design in terms of a constrained minimization of an objective function $f(\mathbf{x}, \mathbf{s})$, given by
\begin{align}
\label{eq:original_problem}
\mathbf{x}_{\textrm{opt}} =& \arg\min_{\mathbf{x}} f(\mathbf{x}, \mathbf{s} ) \ \ \textrm{    s.t. } \mathbf{x} \in   \mathcal{P}^{M} \textrm{.}    
\end{align}
A lower bound on the optimal value for the objective function in \eqref{eq:original_problem} can be obtained by relaxing the problem, e.g., as done in \eqref{eq:lower_bound}.  
The solution of the relaxed problem is termed $\mathbf{x}_{\textrm{lb}}$.
 
An upper bound on the problem \eqref{eq:original_problem} is given by any valid solution which fulfills the restriction that the entries of the vector $\mathbf{x}$ are taken from the discrete set $\mathcal{P}^{M}$. The vector of the discrete set which has a minimum Euclidean distance to the optimal vector of the continuous problem is often suitable for computing an upper bound. In this regard, we denote the smallest known upper bound by $\check{ f } \geq  f(\mathbf{x}_{\textrm{opt}})$, where $\mathbf{x}_{\textrm{opt}}$ is the solution of \eqref{eq:original_problem}.
Now we consider that $d$ entries of $\mathbf{x}$ are fixed and taken out of the discrete set.
The precoding vector is then given by $\mathbf{x}=[\mathbf{x}_1^T, \mathbf{x}_2^T ]^T$, with $\mathbf{x}_1 \in \mathcal{P}^d $.
Based on that, a subproblem can be formulated by
\begin{align}
\label{eq:lb_subproblem}
\mathbf{x}_{2,\textrm{lb}} =& \arg\min_{\mathbf{x}_2} f(\mathbf{x}_2, \mathbf{x}_1, \mathbf{s} ) \\
&\textrm{s.t. }   \left| \mathrm{Re} \left\{   [\mathbf{x}_2]_m  \right\}\right| \leq  1/\sqrt{2M},     \notag \\
& \ \ \ \   \left| \mathrm{Im} \left\{   [\mathbf{x}_2]_m  \right\}\right| \leq  1/\sqrt{2M}, , \ \   \textrm{for }m=1\ldots{M-d}\textrm{.}  \notag  
 \end{align}
If the optimal value of \eqref{eq:lb_subproblem} is larger than a known upper bound $\check{ f }$ on the solution of \eqref{eq:original_problem} the fixed vector $\mathbf{x}_{1}$ and all its evolutions can be excluded from the possible candidates. The branch-and-bound method is efficient when there are many exclusions and the bounds can be computed with relatively low complexity.

\subsection{Proposed 1-Bit Branch-and-Bound Algorithm (1-Bit B\&B)}
\label{sec:bb_algorithm_design}
In this section a branch-and-bound algorithm is proposed which solves \eqref{eq:optimization_problem_precoder} by employing \eqref{eq:lower_bound}, where a real-valued description is used.
The real-valued representations of the precoding vector and the transmit symbol vector are given by\looseness-1
\begin{align}								
\mathbf{x}_{\textrm{r}}					=			\begin{bmatrix} \mathrm{Re} \left\{\mathbf{x}\right\} \\  \mathrm{Im} \left\{\mathbf{x}\right\} \end{bmatrix} \textrm{,} \qquad
\mathbf{s}_{\textrm{r}}					=			\begin{bmatrix} \mathrm{Re} \left\{\mathbf{s}\right\} \\  \mathrm{Im} \left\{\mathbf{s}\right\} \end{bmatrix}  \textrm{,} 
\end{align}
and the real-valued notation of the channel matrix is given by
\begin{align}
\mathbf{H}_{\textrm{r}}    &=       \begin{bmatrix}    \mathrm{Re} \left\{  \mathbf{H}  \right\}  &  -\mathrm{Im} \left\{  \mathbf{H}  \right\} \\
								\mathrm{Im} \left\{  \mathbf{H}  \right\} &  \ \mathrm{Re} \left\{  \mathbf{H}  \right\} 
								\end{bmatrix} \textrm{,}
\end{align}
such that the real-valued noiseless received vector is $\mathbf{r}_{\textrm{r}} =\mathbf{H}_{\textrm{r}} \mathbf{x}_{\textrm{r}}$ and the received vector is $\mathbf{y}_{\textrm{r}}=[\mathrm{Re}\left\{\mathbf{y}\right\}^T \mathrm{Im}\left\{\mathbf{y}\right\}^T ]^T$ and equivalently for the $k$th user the specific notation $\mathbf{H}_{\textrm{r},k}$, $\mathbf{y}_{\textrm{r},k}$ and $\mathbf{s}_{\textrm{r},k}$.
With the vector of variables of the optimization problem described by $\mathbf{v}=[\mathbf{x}_{\textrm{r}}^T, \epsilon]^T$ the optimization problem can be written as:
\begin{align}
\label{eq:v_real}
\mathbf{v}_{\text{opt}}=& \arg\min_{\mathbf{v}}  \boldsymbol{a}^T \mathbf{v} \\
& \textrm{s.t. } \mathbf{A} \mathbf{v} \geq  \boldsymbol{0}_{2 K L},  \notag \\
&     \left|\left[\mathbf{v}\right]_{m}\right|   \in \mathcal{P}_{\mathrm{r}},  \ \  \textrm{for } m=1,\ldots,2M     \textrm{,} \notag
\end{align}
where $\boldsymbol{a}=[\boldsymbol{0}_{2M}^T,-1]^T$, 
$\mathbf{A}= \begin{bmatrix}    \mathrm{diag} \left\{ \mathbf{s}_{\textrm{r}} \right\} \mathbf{H}_{\textrm{r}}, -\boldsymbol{1}_{K L}     \end{bmatrix}$.
In the branch-and-bound method subproblems are solved due to $\mathbf{v}=\left[\mathbf{v}_1^T,\mathbf{v}_2^T   \right]^T$, where $\mathbf{v}_1$ is a fixed vector of length $d$, taken from the discrete set $\mathcal{P}_{\mathrm{r}}^d$ with $\mathcal{P}_{\mathrm{r}}:=\left\{ 1/\sqrt{2M}, -1/\sqrt{2M}  \right\}$.
Accordingly the matrix of the first inequality in \eqref{eq:v_real} is expressed as $\mathbf{A}=\left[\mathbf{A}_1,  \mathbf{A}_2   \right]$, where $\mathbf{A}_1$ contains the first $d$ columns of $\mathbf{A}$.
The lower bound on the subproblem associated with the vector $\mathbf{v}_2$ is given by \looseness-1
\begin{align}
\label{eq:bb_optimzation_problem}
{\mathbf{v}_{2,\textrm{lb}}}=& \arg\min_{\mathbf{v}_2}  \boldsymbol{a}_2^T \mathbf{v}_2 \\
& \textrm{s.t. } \mathbf{A}_2 \mathbf{v}_2 \geq  \mathbf{b}    \notag \\
&     \left|\left[\mathbf{v}\right]_{m}\right|   \leq  c  \ \  \textrm{for all } m=1,\ldots,2M-d     \textrm{,} \notag
\end{align}   
where $\boldsymbol{a}_2=\left[ \boldsymbol{0}_{2M-d}^T,-1 \right]^T$ and $\mathbf{b}=-\mathbf{A}_1 \mathbf{v}_1$. 
The aim is to reduce the number of candidates by excluding individual vectors $\mathbf{v}_1$ based on the lower bounds obtained by \eqref{eq:bb_optimzation_problem} and upper-bounding of \eqref{eq:optimization_problem_precoder}, such that the solution finally can be obtained by an exhaustive search.
For the considered problem \eqref{eq:optimization_problem_precoder} a breadth-first strategy is proposed, where the individual steps are described in Algorithm~\ref{alg:BB_Precoding}. The subproblem \eqref{eq:bb_optimzation_problem} is an LP that can be solved by active set methods, which can take advantage of initialization vectors near the optimum. This can be practically exploited by the branch-and-bound strategy where a series of similar problems have to be solved.
\begin{algorithm}
  \caption{Proposed 1-Bit B\&B Precoding for solving \eqref{eq:optimization_problem_precoder}}
	\label{alg:BB_Precoding}
  \begin{algorithmic}    
	\Initialization{}
		\vspace{-1.25em}
	\State{Given the channel $\mathbf{H}$ and transmit symbols $\mathbf{s}$ compute a valid upper bound $\check{f}$ on the problem in \eqref{eq:optimization_problem_precoder}, e.g., by solving \eqref{eq:lower_bound} followed by a mapping to the closest precoding vector $\mathbf{x}_{\textrm{r}} \in \mathcal{P}_{\textrm{r}}^{2M}$}
	\vspace{2mm}
	\State{Define the first level ($d=1$) of the tree by $\mathcal{G}_{d}:=\mathcal{P}_{\textrm{r}}$}
	\vspace{2mm}	
	\For{$d=1:2M-1$}
	\State{ Partition  $\mathcal{G}_{d}$ in $\mathbf{x}_{1,1},\ldots,\mathbf{x}_{1,\left|\mathcal{G}_{d}\right|}$ }  
	
	  \For{$i=1:\left| \mathcal{G}_{d} \right|$}
		\vspace{2mm}
		\State{Given $\mathbf{x}_{1,i}$ and $\mathbf{s}_{\textrm{r}}$ solve $\mathbf{v}_{2,\textrm{lb}}$ from \eqref{eq:bb_optimzation_problem}  }
		\State{Determine $\epsilon=\left[\mathbf{v}_{2,\textrm{lb}}\right]_{2M-d+1}$}
		\State{Compute the lower bound:  $\mathrm{lb}(\mathbf{x}_{1,i}):=  -\epsilon $;}
		\vspace{2mm}
		\State{Map $\mathbf{x}_{2,\mathrm{lb}}$ to the discrete solution with the closest} 
	  \State{Euclidean distance:}
		\State{$\check{\mathbf{x}}_2(\mathbf{x}_{2,\mathrm{lb}}) \in \mathcal{P}_{\textrm{r}}^{2M-d} $}
		\State{Using $\check{\mathbf{x}}_2$ find the smallest (inverted) distance to the}
		\State{decision threshold:}	
		\begin{align*}
		\mathrm{ub}(\mathbf{x}_{1,i}) := \max_{k}  \left[ -\mathrm{diag} \left\{ \mathbf{s}_{\textrm{r}} \right\} \mathbf{H}_{\textrm{r}}  
					        \begin{bmatrix} \mathbf{x}_{1,i} \\   \check{\mathbf{x}}_2   \end{bmatrix}      \right]_k
									\end{align*}  
									\State{Update the best upper bound with:}
		\State{$\check{f} =\min\left( \check{f}, \mathrm{ub}(\mathbf{x}_{1,i})  \right)    $}
	\EndFor
	\State{Build a reduced set by comparing lower bounds with}
	\State{the upper bound}
	\State{$\mathcal{G}_{d}^{\prime}:=\left\{  \mathbf{x}_{1,i} \vert \mathrm{lb}(\mathbf{x}_{1,i})  \leq  \check{f}       , i=1,\ldots,  \left|\mathcal{G}_{d}\right|  \right\} $}
	\vspace{2mm}
	\State{Define the set for the next level in the tree}
	\State{$\mathcal{G}_{d+1}:=\mathcal{G}_{d}^{\prime} \times \mathcal{P}_{\textrm{r}}$}
	\EndFor
	\State{ Partition  $\mathcal{G}_{d+1}$ in $\mathbf{x}_{1,1},\ldots,\mathbf{x}_{1,\left|\mathcal{G}_{d+1}\right|}$ }	
	  \State{  \begin{align*}
		\epsilon(\mathbf{x}_{1,i}) := \min_{k}   \left[  \mathrm{diag} \left\{ \mathbf{s}_{\textrm{r}} \right\} \mathbf{H}_{\textrm{r}}  
					         \mathbf{x}_{1,i}    \right]_k    \end{align*}}	
\State{The global solution is 
\begin{align*}
 \mathbf{x}_{\textrm{opt}} = \mathrm{arg} \max_{\mathbf{x}_{1,i} \in \mathcal{G}_{d+1}} \epsilon(\mathbf{x}_{1,i})  		
\end{align*}}
\end{algorithmic}
\end{algorithm}
Computing the precoding vector in each time instance exceeds the computational capacities in most applications. However, for small arrays and a large coherence time of the channel, a look-up-table can store the set of precoding vectors as suggested in \cite{Jedda_2016}. Considering the symmetries of the constellation $4^{K L-1}$ precoding vectors need to be computed and stored.

\section{Numerical Results}
\label{sec:numerical_results}
\begin{figure}
\begin{center}
\includegraphics{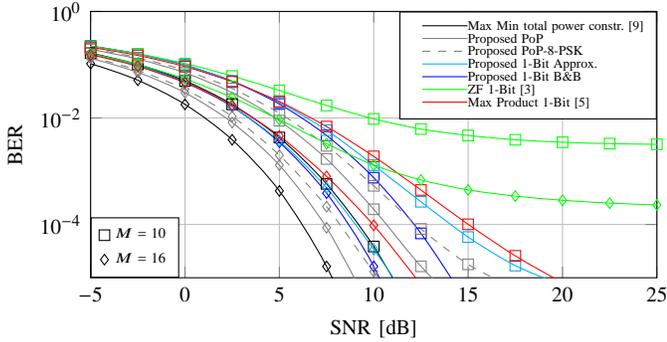}
\captionsetup{justification=centering}
\caption{Uncoded BER versus $\mathrm{SNR}$, $K L=2$, colored curves with 1-Bit DACs } 
\label{fig:BER_1}       
\vspace{-1em}
\end{center}
\end{figure}

\begin{figure}
\begin{center}
\includegraphics{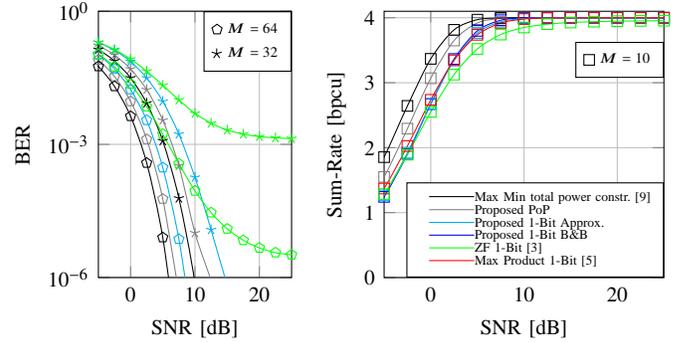}
\captionsetup{justification=centering}
\caption{Uncoded BER (left), $K L=5$; Sum Rate (right), $M=10$, $K L=2$} 
\label{fig:BER_2}       
\vspace{-1em}
\end{center}
\end{figure}

For the numerical evaluation of the uncoded BER and the sum-rate the signal-to-noise ratio is defined by
\begin{align}
\mathrm{SNR}=\frac{\mathrm{E}\left\{ E_{\textrm{Tx}}  \right\}   }{N_0} =\frac{ \left\|\mathbf{x}\right\|^2_2 }{\sigma_n^2} \textrm{,}
\end{align}
where $N_0$ is the noise power density.
We have compared our algorithms with the state-of-the-art precoders with 1-Bit DAC \cite{Saxena_2016} and \cite{Jedda_2016} and the high resolution precoder \cite{Mo_2015} whose performance is slightly better as only its total power is constrained.
Fig.~\ref{fig:BER_1} shows the BER performance of a scenario with two users with a single antenna and different sizes of the array at the transmitter.
As known from the literature the ZF approach \cite{Saxena_2016} shows an error floor which decreases with an increasing number of BS antennas. The proposed 1-Bit methods show a superior performance in terms of BER in comparison to the existing methods using 1-Bit DACs for $M=10$ (and below) and $K L=2$.
For $M=16$ (and above) the approximation, by mapping the solution of \eqref{eq:lower_bound} to the discrete vector with smallest Euclidean distance, and the exact solution of \eqref{eq:optimization_problem_precoder} have almost identical BER.
The optimal utilization of the 1-bit DACs shows a loss of not more than \si{2}{dB} in comparison to the proposed PoP (high resolution), which shows only marginal performance loss when considering an 8-PSK like phase quantization at the transmitter, e.g., by using 2-Bit DACs. 
The LHS of Fig.~\ref{fig:BER_2} shows the performance for larger arrays and more users with $K L=5$.
With the quantization at the receivers the channel output is discrete, such that the sum-rate corresponds to the sum over the mutual information of $K$ discrete memoryless channels $\sum_{k=1}^{K} I_k $, where 
\begin{align}
I_k =  \sum_{\boldsymbol{s}_k, \boldsymbol{y}_k }  P(\boldsymbol{y}_k \vert \boldsymbol{s}_k)P(\boldsymbol{s}_k)  \log_2  \frac{ P(\boldsymbol{y}_k \vert \boldsymbol{s}_k)P(\boldsymbol{s}_k)  }{ P(\boldsymbol{s}_k) P(\boldsymbol{y}_k) }~(\textrm{bpcu}) \textrm{,}
\label{eq:sumRate}
\end{align} 
where $P(\boldsymbol{s}_k)=1/4^{L}$, $P(\boldsymbol{y}_k \vert \boldsymbol{s}_k)= P(\boldsymbol{s}^{\prime}) \sum_{\boldsymbol{s}^{\prime}}  \prod_{i=1}^{2L}  1/2~ \mathrm{erfc}( - [\boldsymbol{y}_{\textrm{r},k}]_i [\mathbf{H}_{\textrm{r},k}\boldsymbol{x}_{\textrm{r}}(\boldsymbol{s}_k,\boldsymbol{s}^{\prime})]_i / \sigma_n )$ with $\boldsymbol{s}^{\prime}=[\boldsymbol{s}_1^T,\ldots,\boldsymbol{s}_{k-1}^T,\boldsymbol{s}_{k+1}^T,\ldots,\boldsymbol{s}_{K}^T]^T$ and $P(\boldsymbol{y}_k)=\sum_{\boldsymbol{s}_k} P(\boldsymbol{y}_k \vert \boldsymbol{s}_k) P(\boldsymbol{s}_k)$.
The RHS of Fig.~\ref{fig:BER_2} shows the sum-rate averaged over varied $\mathbf{H}$, where the proposed methods are gainful in medium $\mathrm{SNR}$.

In what follows, we show the benefit over the exhaustive search, which yields the same solution, by a pessimistic complexity estimation.
Assuming that the subproblems \eqref{eq:bb_optimzation_problem}, which are LPs, are solved by interior point methods (IPMs), the complexity of a subproblem is according to a widely accepted estimate in the community \cite{boyd_2004} in the order of $\mathcal{O}(n^{3.5})$, with $n\leq2M+1$.
Moreover, the average number of visited branches for the proposed branch-and-bound approach and the setting in Fig.~\ref{fig:BER_1} is shown in Table~\ref{tab:efficiency}, which is based on these numbers in the order of $\frac{3}{5}(2 M)^{2.5} $.
With this, the proposed method has polynomial complexity whereas the exhaustive search has exponential complexity and ZF precoders have a cost of $\mathcal{O}(n^{3})$, which also serves as a tight lower bound for the complexity of \cite{Jedda_2016} in which ZF is used for initialization. 
In our case active set methods (ASMs) are practically more efficient in comparison to IPMs because warm starting is possible, which can be exploited in branch-and-bound schemes with a series of similar problems.
The drawback of ASMs, namely its worst-case number of iterations is only theoretical because of the discrepancy between practice and worst-case behavior \cite{Borgwardt_1987}.

\section{Conclusions}
We have proposed an approach for optimal precoding for DACs with 1-bit resolution.
The design criterion describes the maximization of the minimum distance to the decision threshold at the receivers. Both the exact solution and the approximation based on the continuous relaxation followed by a mapping to the discrete set yield outstanding performance in terms of uncoded bit error rate, which can serve as a benchmark for designers working on 1-bit precoding techniques. The simulation results show that the loss brought by 1-Bit DACs is less than \si{2}{dB}.
\begin{table}
 \vspace{0pt}
\begin{center}
\captionsetup{justification=centering,font=scriptsize}
\caption{Efficiency of the proposed branch-and-bound approach}
\vspace{-0.5em}
\includegraphics{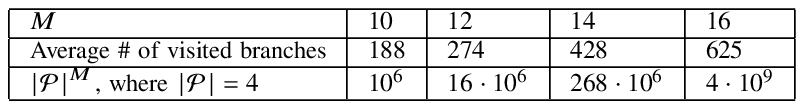}
			\label{tab:efficiency}
		\end{center}
		\vspace{-2em}
		\end{table}

\bibliographystyle{IEEEtran}

\end{document}